\documentclass{rsproca_new}

\usepackage{amsmath}  
\usepackage{amsfonts} 
\usepackage{graphicx} 
\usepackage{epigraph}
\usepackage{cancel}
\usepackage{mathtools}
\usepackage[normalem]{ulem}
\usepackage{commath}
\usepackage{float}
\usepackage{multimedia}
\usepackage{tabularx}
\usepackage{xcolor}
\usepackage{hyperref}
\hypersetup{
    colorlinks=true,
    linkcolor=red,
    citecolor=blue,      
    urlcolor=blue,
    breaklinks=true
}

\makeatletter
\newcommand{\vast}{\bBigg@{4}}
\newcommand{\Vast}{\bBigg@{5}}
\makeatother

\begin{document}

\title{Gamow's Cyclist: A New Look at Relativistic Measurements for a Binocular Observer}

\author{E. A. Cryer-Jenkins$^1$ and P. D. Stevenson$^1$}
\address{$^1$Department of Physics, University of Surrey, Guildford, GU2 7XH, United Kingdom}
\corres{P. D Stevenson\\
\email{p.stevenson@surrey.ac.uk}}

\subject{Special Relativity}

\date{20th May 2019}

\begin{abstract}
The visualisation of objects moving at relativistic speeds has been a popular topic of study since Special Relativity's inception.  While the standard exposition of the theory describes certain shape-changing effects, such as the Lorentz-contraction, it makes no mention of how an extended object would appear in a snapshot or how apparent distortions could be used for measurement. Previous work on the subject has derived the apparent form of an object, often making mention of George Gamow's relativistic cyclist thought experiment. Here, a rigorous re-analysis of the cyclist, this time in 3-dimensions, is undertaken for a binocular observer, accounting for both the distortion in apparent position and the relativistic colour and intensity shifts undergone by a fast moving object. A methodology for analysing binocular relativistic data is then introduced, allowing the fitting of experimental readings of an object's apparent position to determine the distance to the object and its velocity. This method is then applied to the simulation of Gamow's cyclist, producing self-consistent results.
\end{abstract}

\begin{fmtext}
\section{Introduction} 
\label{intro}

In 1938, George Gamow envisioned a beautiful thought experiment in his book \textit{Mr Tompkins' Adventures in Wonderland} \cite{Gamow1958}: in it, the titular hero is transported to a strange world in which the speed of light is only slightly faster than that of a bicycle and he sees a passing cyclist to be Lorentz contracted, in apparent agreement with Einstein's Theory of Special Relativity. This view, that a moving extended object would appear ``simply" contracted, was widely held in the first half of the 20th century with Einstein and Lorentz both promulgating the visibility, and the ability to be photographed, of relativistic length contraction \cite{Terrell1959}.

\end{fmtext}

\maketitle 

Despite an early attempt to correct this misconception by Anton Lampa in 1924 \cite{Lampa1924}, it would not be until the late 1950s when two papers, one by Roger Penrose \cite{Penrose1959} and another by James Terrell \cite{Terrell1959} (after whom this family of visual aberrations is named), that the actual visual appearance of relativistic objects became widely acknowledged.

Images are generated by photons arriving at an observer simultaneously, not by those simultaneously emitted by an object. As light has both a constant and finite speed, what one really sees is an extended object made up of a patchwork of itself at different times. Terrell's conclusion was a simple one: a quickly moving object appears rotated with an increasing proportion of its rear face becoming visible at greater speeds; an elegant solution that quickly found its way into the works of others and into the educational literature \cite{Weisskopf1960,Peres1987,Resnick1968}. In the years following, re-analyses of Terrell's work (most significantly by Mathews and Lakshmanan \cite{ML1972}) painted a picture of a more complex set of distortions which were better understood as a non-linear shear and extension/contraction parallel to the direction of movement. Mathews and Lakshmanan's re-analysis was swiftly joined by a vast array of literature expounding the various phenomenological consequences of this treatment \cite{Boas1961,ScottVanDriel1970,Sufferen1988,Scott1965,Sheldon1988,McKinley1980,Nowojewski2004,Manoukian2002,Savage,Dreissler2005}, some pieces citing Gamow's cyclist as a metric with which to consider the extent of these deformations \cite{Boas1961,Nowojewski2005}. These results describe what Mr Tompkins would \textit{photograph} in these relativistic situations (assuming the photographic device able to detect photons of any wavelength) but not, as Mr Tompkins is a human and not a camera, what he would actually \textit{see}.

In this work, we have re-analysed relativistic visualisation from the perspective of a binocular observer, introducing an analytical time delay between the ``eyes" of such an observer and applying it to the analytical transformations of an object's actual location to its apparent one as well as the relativistic Doppler and intensity shifts of a body isotropically emitting radiation in its rest frame. This treatment is then applied to a 3-dimensional simulation of Gamow's cyclist, comprised of spheres and cylinders, thus giving a more complete picture of what Mr Tompkins might see. 

After reviewing the analytical results of the visual relativitic effects, we produce a spherical polar method, using binocular observers, of fitting experimental position data from moving objects to determine their actual velocity and distance to the observer, allowing corrections to be made to multi-aperture relativistic photographs and other quantities, such as emission intensities and spectra, to be determined. Finally, it is suggested that these methods could be integrated within a relativistic probe, such as one proposed in the work of Christian and Loeb \cite{Christian2017}, or a synthesised, ground-based aperture that would allow the probing of exoplanets and other astronomical bodies.

\section{Mathematics of Relativistic Distortions}
\label{mathsreldist}

In this section, we review - in accordance with Einstein's Postulates of Special Relativity \cite{Einstein1905} - the equations that translate various physical quantities from their actualities to their apparencies, already present in literature \cite{ML1972}. We make use of the trigonometrical relations that hold in Euclidean, flat space and this result is thus valid only in the case that strong gravitational fields are not present. The terms $\beta=v/c$ and $\gamma=(1-\beta^2)^{-1/2}$ represent their standard quantities in Special Relativity.

\subsection{Apparent Positions}
\label{apparentpositions}

We follow Mathews and Lakshmanan \cite{ML1972} in using \textit{objective} to refer to properties of the (moving) object in a frame in which it is moving that are due to ``purely relativistic effects'' that could be determined by some contact method without the intervention of light signals.  The \textit{apparent} quantities then take into account the communication of properties of the object by light travelling to the observer which is perceived at a single point in spacetime.

For a point $P$, travelling parallel to the $x$-axis in the positive $x$-direction with coordinates $P(vt',y,z)$, the apparent position, $x'$, is given by
\begin{equation}
    \label{eq:apparentx}
    x' = \gamma^2(x_0 +\beta cT) - \frac{\gamma}{c}\sqrt{\gamma^2(x_0+\beta cT)^2+y_0^2+z_0^2} 
\end{equation}
for an observer on the $x$-axis where $t'$ is the time at which a photon is emitted and $T$ is the time at which the same photon is received at the observer. $x_0$, $y_0$ and $z_0$ are the $t'=0$ objective coordinates of the point \emph{in the observer's coordinate frame} and $v$ is the point's velocity. For an observer at the origin, we therefore have that $x_0=0$, $y_0=y$ and $z_0=z$.

The other apparent coordinates, $y'$ and $z'$, retain their objective values, $y_0$ and $z_0$ respectively.  Here, we will only consider points with $x$-coordinate given by $$x(t')= vt'$$ so that $x(0)=0$.

\subsection{Relativistic Radiative Effects}
\label{reldopplershift}

Along with visual distortions, we also consider the colour change of the object in question, both in the interest of phenomenologically examining what Mr Tompkins might see and also for scientific measurement, as photographs of relativistic objects must be corrected for Doppler and luminosity shifts as a result of the motion of a radiating source\cite{Oliver1964,Rees1966,Kraus2000}.  Maintaining the previous configuration of the point, $P$, the relativistic Doppler shift between the received wavelength, $\lambda'_r$, at the observer and source wavelength, $\lambda_S$, emitted by $P$ is given by
\begin{equation}
    \frac{1}{\lambda'_r} = \gamma\bigg(1-\frac{\gamma\beta(x_0+\beta cT)}{\sqrt{\gamma^2(x_0+\beta cT)^2+y_0^2+z_0^2}} \bigg)\frac{1}{\lambda_s}
\end{equation}

Similarly, we find that the intensity at the observer, $I'$, for an isotropically (in its own rest frame) emitting source with rest intensity, $I_0$, is given by
\begin{equation}
   I' = \gamma^2\bigg(1-\frac{\gamma\beta(x_0+\beta cT)}{\sqrt{\gamma^2(x_0+\beta cT)^2+y_0^2+z_0^2}} \bigg)^2 I_0
\end{equation}


\section{Measurements from Binocular Distortion}
\label{measuredistortions}

In this section, we introduce formalism with respect to binocular observations; herein, we define two types of observers:
\begin{enumerate}
    \item[\textbf{Class 1}] An observer with a single aperture such as a camera
    \item[\textbf{Class 2}] An observer with two apertures, capable of depth perception generated by visual parallax such as a human 
\end{enumerate}

All apertures in these definitions are considered to have an infinitesimal exposure time and perfect focus.
\\
\\
Works by Boas\cite{Boas1961} and Nowojewski\cite{Nowojewski2005} have gone some way towards realising the actual form of Gamow's cyclist but, as with the other considerations made before, they deal with the aberrations presented to a Class 1 observer. As such, Mr Tompkins would not see the bicycle as described by Gamow, Nowojewski, Boas or others \cite{MullerBoblest2014,Hickey1979} due to him being a Class 2 observer.  We also choose to account for colouration and intensity shifts \cite{Kraus2000} to provide a more complete picture of our relativistic cyclist, including the difference between distortions presented to each aperture providing a method of determining the distance from and speed of the bicycle.

\subsection{Time Delay between Observers}

\begin{figure}[H]
    \centering
    \includegraphics[width=0.6\textwidth]{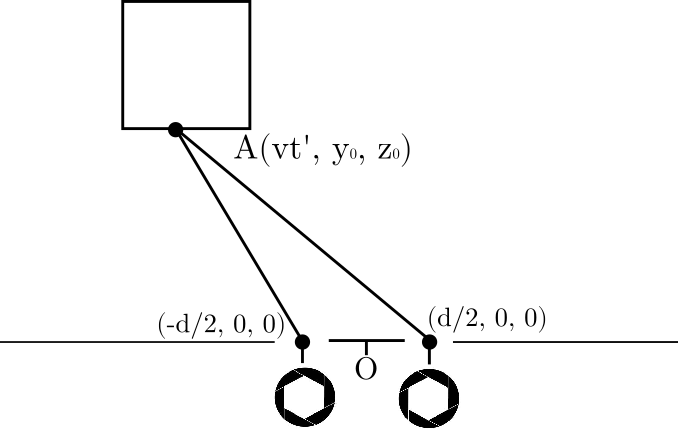}
    \caption{Construction used when considering the time difference between two apertures, evenly spaced around the origin, $O$, a distance $d$ apart. After arriving at the $x$-axis, photons experience no additional path difference before reaching an aperture.}
    \label{fig:twoeyes}
\end{figure}

A Class 2 observer's two apertures will perceive the same object differently at different points in its trajectory as the distance between them is not negligible for relativistically moving bodies. As such, it is useful to introduce a time difference between apertures so that one equation can be used with identical values apart from a time difference term, $\Delta T$, between the \emph{primary} and \emph{secondary} apertures. Here, we consider a Class 2 observer with apertures equally spaced a distance $\frac{d}{2}$ either side of the origin, the ``left" aperture at $x=-\frac{d}{2}$ being the primary. The object, as before, is made of points with coordinates $(vt', y_0, z_0)$, travelling with velocity $v$; we see this construction presented in Figure \ref{fig:twoeyes}.

To find this time difference, we consider the path difference from Point A to each aperture. Defining $v$ as positive, Point A moves from left to right and a ``newer" realisation of the object will initially be presented to the left aperture until it passes the point of equidistance at $t'=0$, after which it is the right aperture realising a ``newer" image. We will therefore define this $\Delta T$ as the time difference added on to the $T$ value \emph{of the left aperture} as it is positive for $t'<0$, motivating its assignment as the primary aperture. We can see this path difference is
\begin{equation}
    \label{eq:timediff}
    c\Delta T = \sqrt{y_0^2+z_0^2+\bigg(vt'-\frac{d}{2}\bigg)^2}-\sqrt{y_0^2+z_0^2+\bigg(vt'+\frac{d}{2}\bigg)^2}
\end{equation}
which provides the expected result; namely that the time difference tends to a constant value when the object is far to either side of the observer and is equal to 0 for $t'=0$, i.e. when the object is equidistant from both apertures.

We then use the trigonometric relation between emission and observation times
\begin{equation}
    c(T-t')=\sqrt{(x_0+vt')^2+y_0^2+z_0^2}
\end{equation}
for $t'=0$ and, as we are considering the left aperture as the primary aperture, $x_0=\frac{d}{2}$, to obtain an expression for the time of apparent equidistance, $T_{t'=0}$ as
\begin{equation}
    \label{eq:intersectiontime}
    T_{t'=0} = \frac{1}{c}\sqrt{\bigg(\frac{d}{2}\bigg)^2+y_0^2+z_0^2}.
\end{equation}
We will see that this is a useful expression in the fitting of visual data as it provides the observer's time at which the object is apparently equidistant from both apertures, allowing the determination of object distance from the fitting of experimental position data.

\subsection{Parameter Fitting}
\label{pf}

\begin{figure}[tbh]
    \centering
    \includegraphics[width=0.9\textwidth]{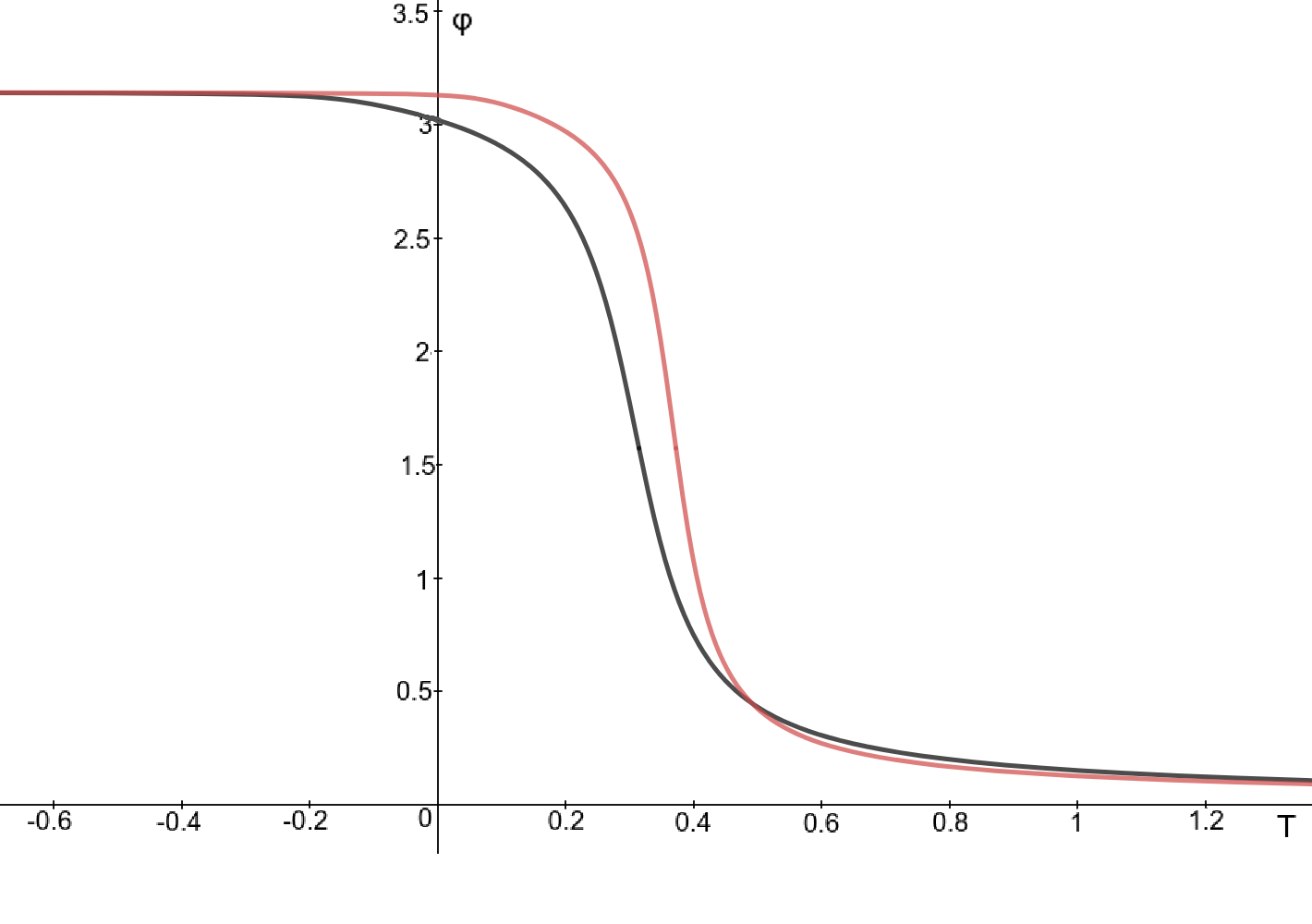}
    \caption{Graph depicting the azimuthal angles, $\phi$, subtended by a point with $y_0=0.5$, $z_0=0.5$ travelling at $\beta=0.9$ (black representing the primary aperture and \textcolor{red}{red} the secondary) for $d=0.3$.}
    \label{fig:Phipos}
\end{figure}

Suppose we can identify a point on the surface of an object approaching the observer with velocity $v$ in the increasing $x$-direction and coordinates $(vt', y_0, z_0)$, the apparent azimuthal angle subtended is given by standard trigonometry

\begin{equation}
    \phi_P=\arctan\bigg(\frac{y_0}{x'(T)}\bigg)
\end{equation}
for the primary aperture, which, we remind, is assigned as being on the left (noting that $\phi$ is defined relative to the positive $x$-axis). The secondary aperture, on the right, will receive, at the same observation time, a ``later" image of the object for $t'<0$, resulting in an apparent azimuthal angle given by
\begin{equation}
    \phi_S=\arctan\bigg(\frac{y_0}{x'(T+\Delta T)}\bigg)
\end{equation}
As we have seen previously, $x'$ is given, in terms of objective coordinates, by
\begin{equation}
    x' = \gamma^2(x_0 +\beta cT) - \frac{\gamma}{c}\sqrt{\gamma^2(x_0+\beta cT)^2+y_0^2+z_0^2}
\end{equation}
We also require an expression for $\Delta T$ in terms of $T$.  We again make use of our expression for $t'$ in terms of objective coordinates and observer time $T$, given by
\begin{equation}
    t' = \frac{\gamma^2}{c}(Tc+x_0\beta) - \frac{\gamma}{c}\sqrt{\gamma^2(x_0+\beta cT)^2+y_0^2+z_0^2}
\end{equation}
which we can substitute into our equation for $\Delta T$ to achieve a non-linear equation for $x'(T+\Delta T)$, the apparent position according to the secondary aperture, in terms only of $T$
\begin{equation}
    \label{eq:righteyeposition}
    x'|_{(T+\Delta T)} = \gamma^2(x_0 +\beta c(T+\Delta T)) - \frac{\gamma}{c}\sqrt{\gamma^2(x_0+\beta c(T+\Delta T))^2+y_0^2+z_0^2} 
\end{equation}
where 
$$\Delta T \equiv \Delta T(x, y_0, z_0, T)$$
\\
\\
The power of expression (\ref{eq:righteyeposition}) is that it allows the determination of relativistic quantities as the differences in polar and azimuthal angles between apertures in Class 2 observers.  These differences have characteristic shapes that can be fitted to determine the velocity of an object and its distance from the observer. We now produce typical plots of $\phi_P$ and $\phi_S$ as well as the difference between them which can be employed for fitting as it provides a second known quantity, $d$. Figure \ref{fig:Phipos} illustrates the azimuthal position of the object for both apertures, indicating a clear cross over as the object appears to ``overtake" itself travelling from the left to the right.
\\
\\
All figures are functions of the observer's time, $T$, and all angles are in radians unless otherwise stated.

\begin{figure}[tbh]
    \centering
    \includegraphics[width=0.9\textwidth]{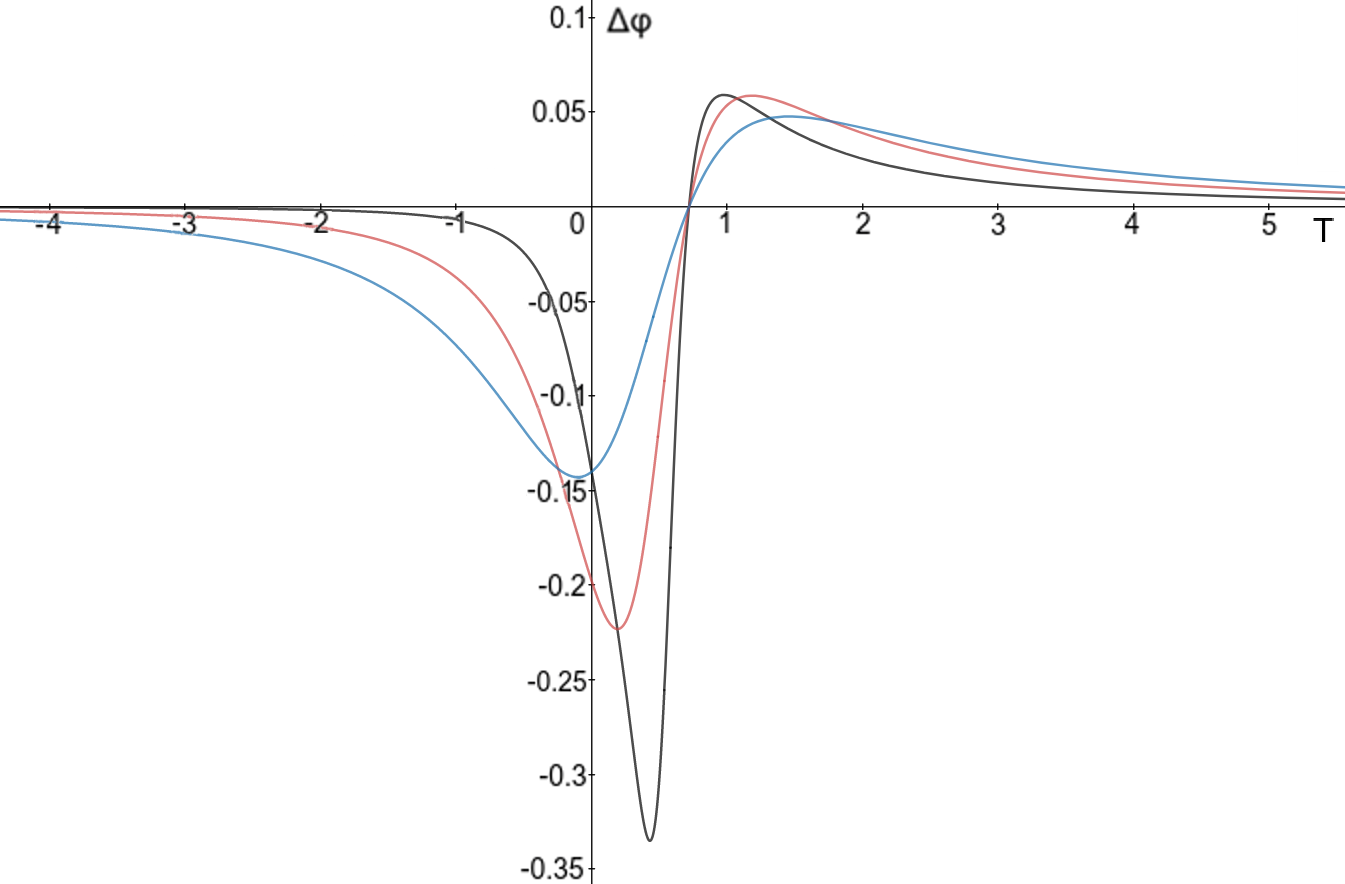}
    \caption{Graph depicting the difference in azimuthal angles, $\Delta\phi$, subtended by a point with $y_0=0.5$, $z_0=0.5$ for an observer with $d=0.3$ and $\beta=$ 0.9, 0.7 and 0.5 for the black, \textcolor{red}{red} and \textcolor{blue}{blue} curves respectively.}
    \label{fig:Phidiff}
\end{figure}

The difference between these apparent azimuthal angles, defined by
\begin{equation}
    \label{eq:phidiff}
    \Delta\phi = \phi_P - \phi_S = \arctan\bigg(\frac{y_0}{x'(T)}\bigg) - \arctan\bigg(\frac{y_0}{x'(T+\Delta T)}\bigg)
\end{equation}
has a characteristic shape, illustrated in Figure \ref{fig:Phidiff}, which is asymmetric due to the difference in apparent velocity for an approaching or receding object. 

Figure \ref{fig:Phidiff} illustrates a maximum negative angular difference of -0.34 rad (for $\beta=0.9$), equivalent to 19.5 degrees and maximum positive difference of 0.06 rad, or 3.4 degrees. As such, these changes are on the order of degrees and are much larger than the smallest resolution of angular measurement devices used in astronomy \cite{Chapman1983}. 
It is also intuitive from the previously defined time difference that, when $\Delta T=0$, the intersection with the $x$-axis in Figure \ref{fig:Phidiff} is given by Equation (\ref{eq:intersectiontime}); as the object is equidistant from both apertures at $t'=0$, it is at coordinates ($\frac{d}{2}$, $y_0$, $z_0$) with respect to the left aperture. Using the analytical data from Figure \ref{fig:Phidiff}, we see (again taking $c=1$ for simplicity)
\begin{equation*}
    T_{t'=0} = \sqrt{(0.15)^2+(0.5)^2+(0.5)^2} = 0.72
\end{equation*}
which is visibly the $x$ intersection in Figure \ref{fig:Phidiff}, providing the distance to the object.
\\
\\
Similar fitting can be carried out for the apparent polar angle (although this is more difficult for smaller/more distant objects), allowing detector sensitivity to be investigated and providing an estimate of the error in object distance and velocity. Here, we define the apparent polar angle again using standard trigonometry
\begin{equation}
    \theta_P = \arctan\left(\frac{z_0}{\sqrt{x'^2(T)+y_0^2}}\right)
\end{equation}
for the primary aperture and
\begin{equation}
    \theta_S= \arctan\left(\frac{z_0}{\sqrt{x'^2(T+\Delta T)+y_0^2}}\right)
\end{equation}
for the secondary.

It should be noted that $\theta$ is defined as the angle subtended by the line joining object and observer and the x-y plane and it thus tends to 0 for $t'\rightarrow\pm\infty$; a closer object also subtends a larger $\theta$.

\begin{figure}[tbh]
    \centering
    \includegraphics[width=1\textwidth]{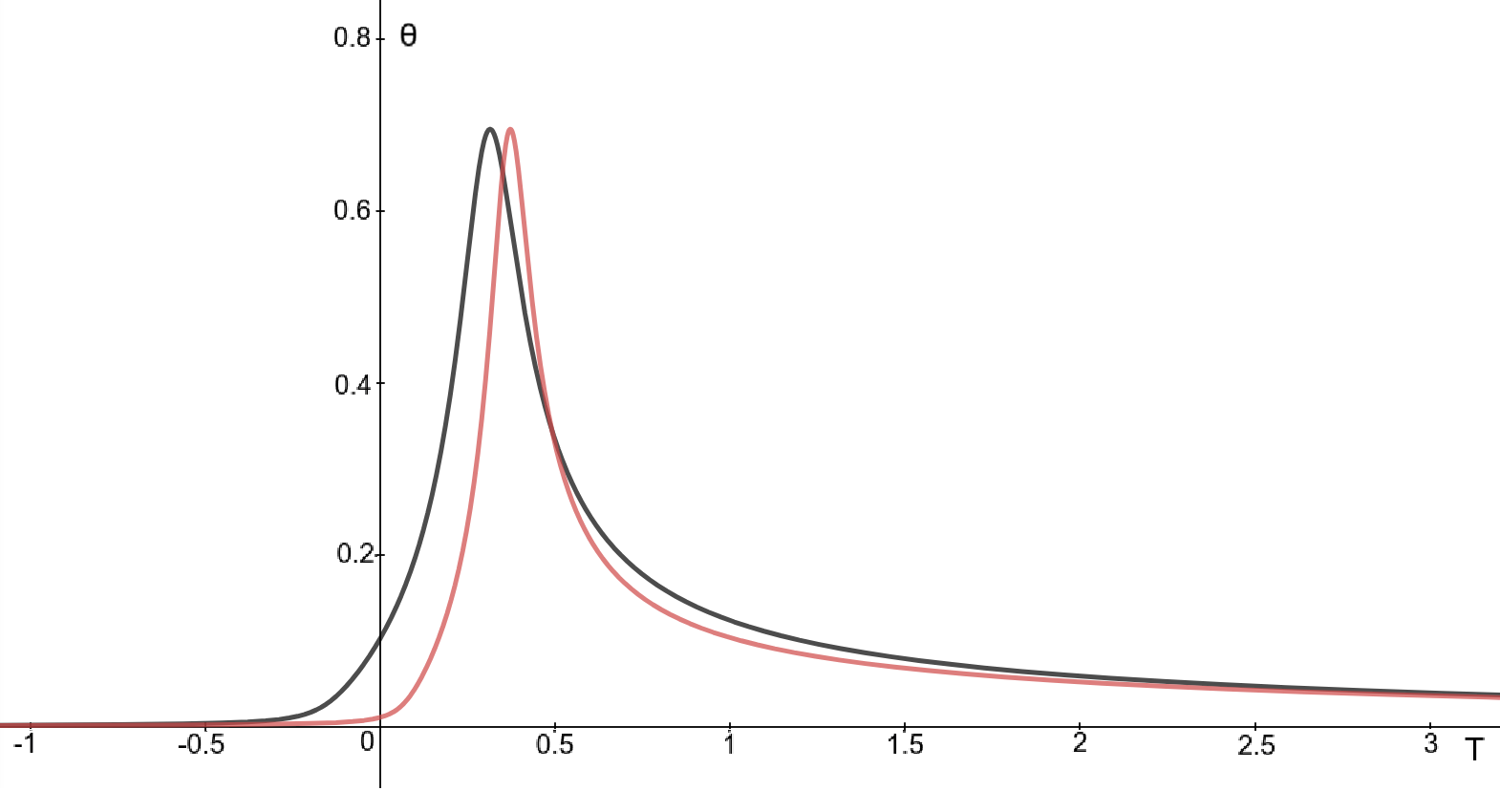}
    \caption{Graph depicting the polar angles, $\theta$ subtended by a point with $y_0=0.5$, $z_0=0.5$ travelling at $\beta=0.9$ (black representing the primary aperture and \textcolor{red}{red} the secondary) for $d=0.3$.}
    \label{fig:Thetapos}
\end{figure}

These quantities have been plotted in Figure \ref{fig:Thetapos}, illustrating the different times at which an object would appear larger for each aperture.

\begin{figure}[H]
    \centering
    \includegraphics[width=1\textwidth]{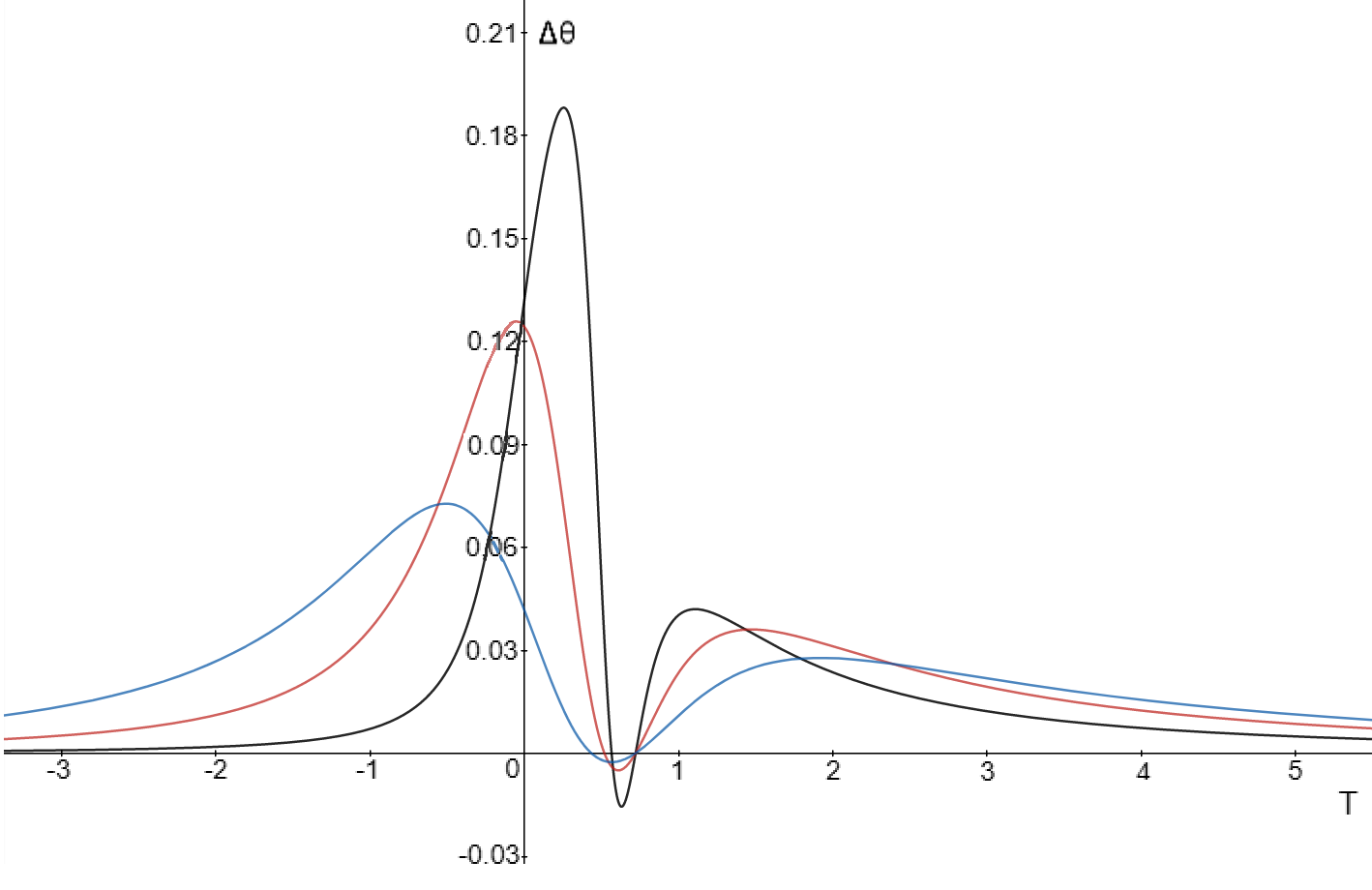}
    \caption{Graph depicting the difference in polar angles, $\Delta\theta$, subtended by a point with $y_0=0.5$, $z_0=0.5$ for an observer with $d=0.3$ and $\beta=$ 0.9, 0.7 and 0.5 for the black, \textcolor{red}{red} and \textcolor{blue}{blue} curves respectively.}
    \label{fig:Thetadiff}
\end{figure}

The difference between these apparent polar angles,
\begin{equation}
    \label{eq:thetadiff}
    \Delta\theta = \theta_P - \theta_S = \arctan\bigg(\frac{z_0}{\sqrt{x'^2(T)+y_0^2}}\bigg) - \arctan\bigg(\frac{z_0}{\sqrt{x'^2(T+\Delta T)+y_0^2}}\bigg),
\end{equation}
also has a characteristic shape, illustrated in Figure \ref{fig:Thetadiff}, which is asymmetric for the same reasons considered previously. 

It also has two intersections with the x-axis which may seem counter-intuitive; however, this too arises from the asymmetry of the construction. The right aperture initially sees an ``older" instance of the object that is further away and thus smaller. As the object passes the left aperture, the right begins to see a larger instance and the angular difference becomes negative. Finally, after the object has just passed the point of equidistance between apertures, the right aperture now receives a ``newer" instance of the object that is further away and thus smaller, reversing the angular difference to positive once more. 

The fitting of the intersection at larger $T$ provides a $\beta$-independent measure of the position, identical to the azimuthal intersection, while the intersection at smaller $T$ is $\beta$-dependent; these points tend to the same time coordinate, $T_{t'=0}$, given by Equation (\ref{eq:intersectiontime}), for larger values of $\beta$ (as in Figure \ref{fig:Thetadiff}) and greater distances as the more extreme instances of the object overtaking itself are smoothed out.
\\
\\
Figure \ref{fig:Thetadiff} illustrates a maximum negative angular difference of -0.015 rad (for $\beta=0.9$), equivalent to 0.86 degrees and maximum positive difference of 0.19 rad, or 10.89 degrees. As noted before, these angles, while for a relatively close object, are sufficiently significant to be readily measured and are orders of magnitude larger than the resolution of current astronomical angle measuring devices.
\\
\\
From fitting Figures \ref{fig:Phidiff} and \ref{fig:Thetadiff} with the aforementioned equations, we can obtain values and combined uncertainties for the distance of the object from the observer and the speed with which the object is moving; these quantities can then be utilised to correct relativistic photographs or emission spectra of moving objects. We will now consider the phenomenological example of Gamow's cyclist, considering both their appearance to a human observer and applying Equations (\ref{eq:phidiff}) and (\ref{eq:thetadiff}) to a point on the bicycle as it moves past, testing the validity of the method.

\section{Gamow's Cyclist}
\label{gamowcyclist}

In this section, we consider the famous and oft-mentioned thought experiment that is Gamow's relativistic cyclist, re-imagining it in 3-dimensions for a Class 2 observer and  applying the methodology set out in the previous section to determine its velocity and distance from the observer.

\subsection{2-Dimensional Appearance}

\begin{figure}[tbh]
    \centering
    \includegraphics[width=0.5\textwidth]{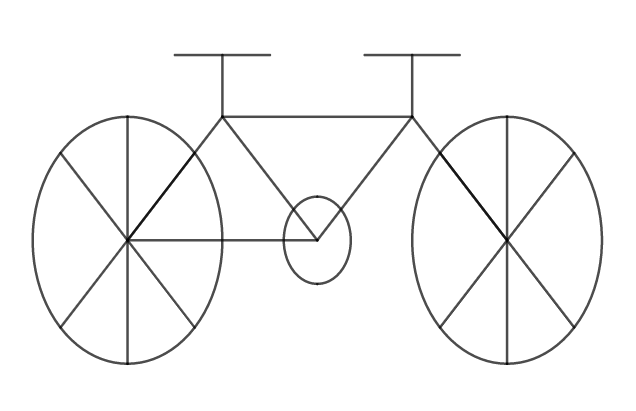}
    \caption{Illustration of a simple bicycle composed of straight lines and circles.}
    \label{fig:bike}
\end{figure}

 First, we consider a simple bicycle composed of straight lines and circles, depicted in Figure \ref{fig:bike}, in order to provide a clear metric with which to measure the distortion of well-known shapes. The transformation of extended objects can be done by considering them as comprising of a continuum of points, each transformed according to Equation (\ref{eq:apparentx}). 

\begin{figure}[tbh]
    \centering
    \includegraphics[width=1\textwidth]{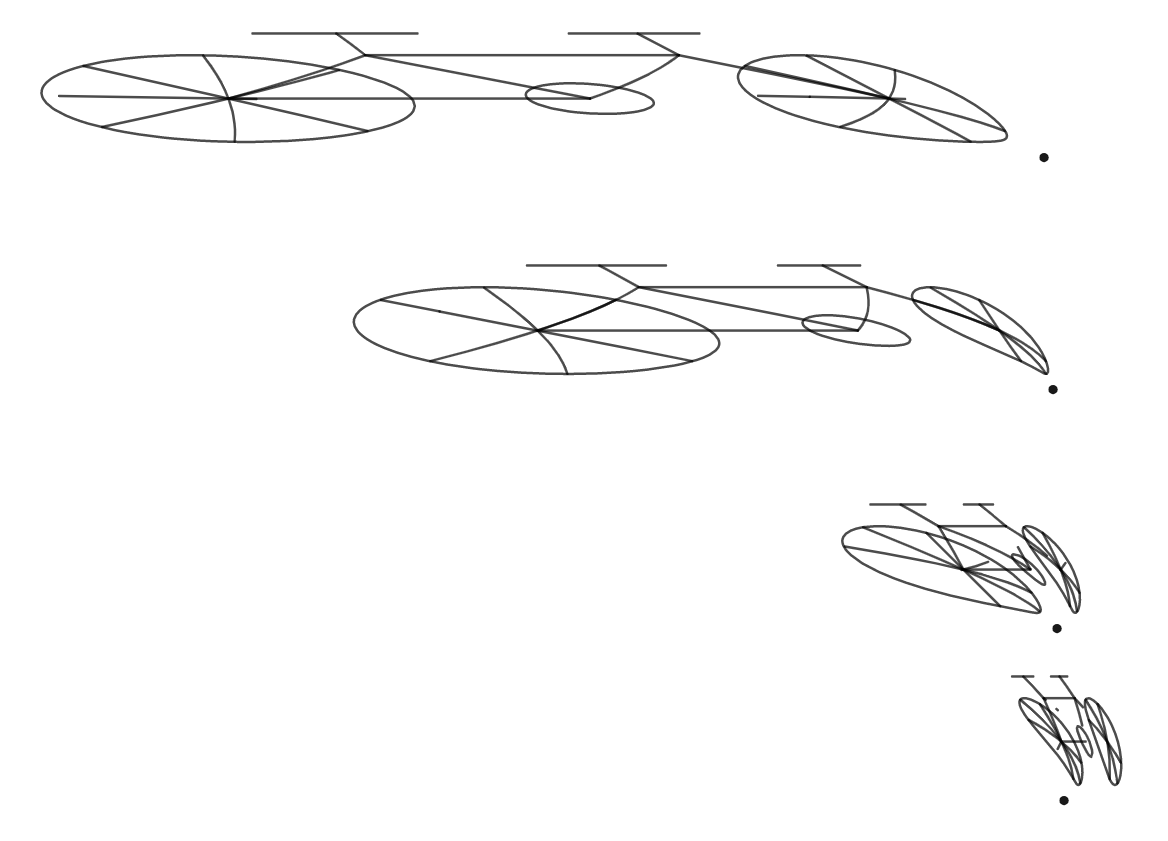}
    \caption{Figure illustrating deformation of a 2D bicycle with $\beta=0.9$ with the observer represented as a black dot.}
    \label{fig:deformedbike}
\end{figure}

Here, we consider this bicycle for a single observer as it travels past, deformed using Mathews and Lakshmanan's transformations in 2 dimensions, not accounting for any kind of intensity or colouration distortion. The visual deformations undergone by the relativistic bicycle are presented in Figure \ref{fig:deformedbike} with the observer represented as a black dot; we see the exact deformations predicted by Mathews and Lakshmanan in 2 dimensions, namely a combination of non-uniform shear and extension/contraction parallel to the direction of movement. As the bicycle approaches the observer (with an apparent velocity greater than $c$ \cite{Burke1991,Terrell1960,Nowojewski2005}), it appears grossly extended with the circular wheels deformed into elongated ellipsoids. As it passes the observer, the wheels and spokes appear concave and the entire frame is contracted (receding now with apparent velocity $\leq \frac{c}{2}$ \cite{Berlet2011}). It can also be noted that all line sections parallel to the direction of movement remain parallel, such as the cross bar of the bicycle, and the height of the bicycle remains unchanged.

\begin{figure}[tbh]
    \centering
    \includegraphics[width=1\textwidth]{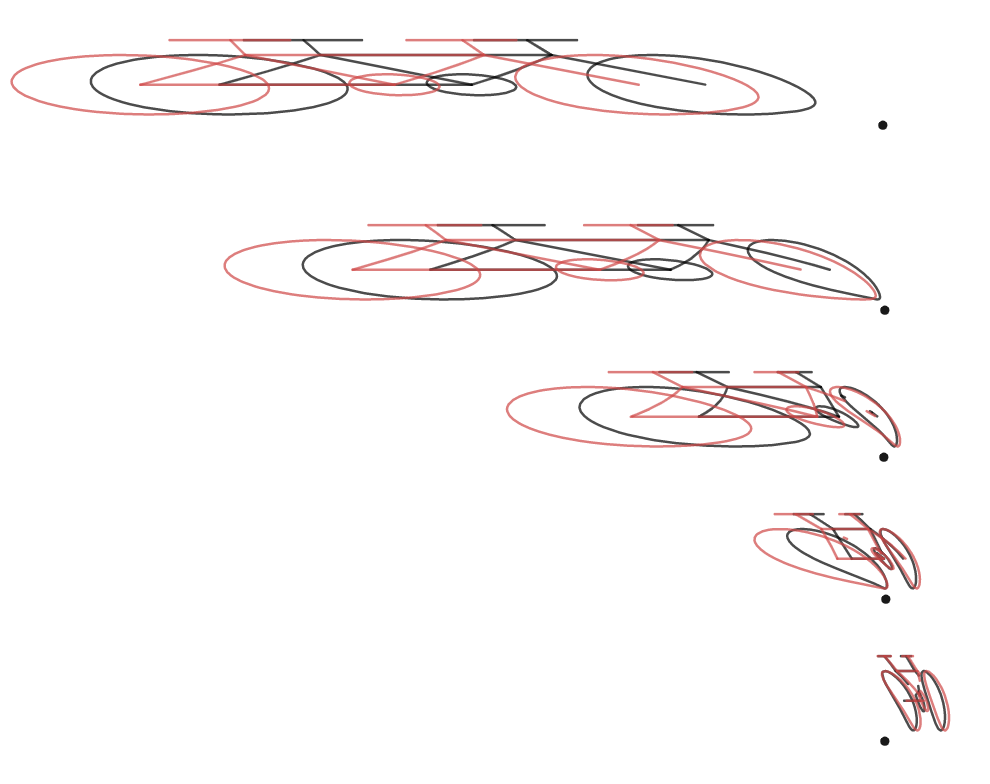}
    \caption{Figure illustrating deformation of a 2D bicycle with $\beta=0.9$ for a Class 2 observer with aperture spacing $d=0.3$. We see the features pointed out in the previous section with different parts of the bicycle catching up with itself at different times. The image received at the primary aperture is in black and at the secondary is in \textcolor{red}{red}.}
    \label{fig:twobikes}
\end{figure}

This construction goes a long way in illustrating the increased elongation and contraction an object experiences before and after passing an observer with ratio $\sqrt{\frac{1+\beta}{1-\beta}}$ before reaching the observer and $\sqrt{\frac{1-\beta}{1+\beta}}$ after passing them. However, there are other considerations to be taken into account for what Mr Tompkins would really see; not only are there colouration shifts, there is also the matter of binocular vision becoming confused with the overlap of contradictory images and Mr Tompkins' perception of 3 dimensions. 
\\
\\
Employing the time difference term, $\Delta T$, from Equation (\ref{eq:timediff}) we can visualise Figure \ref{fig:deformedbike} but for a Class 2 observer with the left aperture again chosen as the primary, providing a good indication of what Mr Tompkins' eyes would actually see, the result displayed in Figure \ref{fig:twobikes}.

As visible in Figure \ref{fig:twobikes}, Mr Tompkins's sensory apparatus, which relies on visual parallax to determine depth \cite{BW2011,Battro1977,Smith1959}, would see apparent fluctuations in distance to the cyclist as the image produced in the secondary aperture (which is denoted by the red bicycle) overtakes that of the left. This disparity between images that so confuses a human is, however, what can be exploited as the basis of a relativistic measuring tool.

\subsection{3-Dimensional Appearance}

\begin{figure}[tbh]
    \centering
    \includegraphics[width=0.7\textwidth]{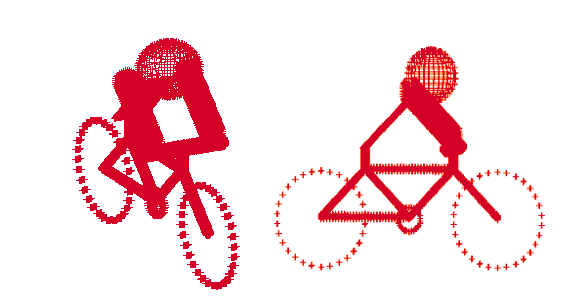}
    \caption{Construction of Gamow's cyclist for which the 3D simulation taking into account colour and intensity was carried out.}
    \label{fig:3Dbike}
\end{figure}

We now construct a simple 3-dimensional bicycle and rider, comprised entirely of tubes and spheres, and displayed in Figure \ref{fig:3Dbike}, which - for simplicity - is entirely comprised of red pixels with wavelength 700nm.

This 3D model was generated from an array of spatial coordinates for a series of pixels, along with emission time $t'$, rest wavelength and intensity of each pixel.  These are then translated into apparent coordinates, observation time $T$, relativistic Doppler wavelength and intensity at the observer through the equations derived previously. The colours and intensities are then transformed and the results animated at 15 frames per second. The crux of this methodology is to allow the transformation of intensities and emission wavelengths back into their rest quantities that enable the analysis of relativistic objects. From Section \ref{measuredistortions}, we have seen that the fitting of binocular profiles, obtained by two (or potentially more to improve the fitting) observers, can be used to determine both the distance to an object and also its velocity.  
\\
\\
In the computer simulation, we track the first point on the wheel (with $t'=0$ coordinates $x_0=0$, $y_0=3$ and $z_0=2$ and velocity $\beta=0.9$ in this simulation), imagining that the observer is trying their best to keep their view on the front end of the bicycle; as such, the entire elongated bicycle does not fit within the visual range of the observer. The distance between apertures is fixed at $d=0.1$. The leftmost image in each frame is the appearance presented to the left aperture; likewise for the rightmost image corresponding to the right aperture. The time at which the forward-most pixel in the front wheel is observed as being equidistant between both apertures is $T_{t'=0}$.
\\
\\
For $T<T_{t'=0}$, we are presented with a frontal view of the bicycle; while the headlight effect relativistically beams the majority of the isotropically emitted radiation forwards, it is Doppler shifted out of the visual spectrum (and into the ~200nm ultraviolet range) and is thus invisible to a human observer (represented as black in the simulation):

\begin{figure}[tbh]
    \centering
    \includegraphics[width=1\textwidth]{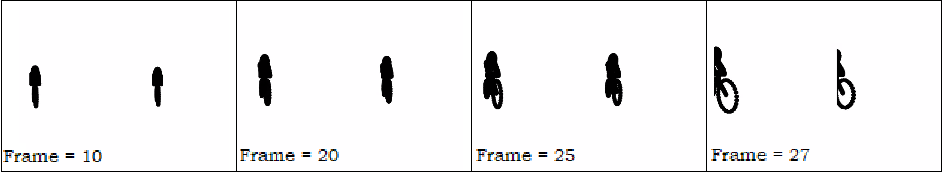}
    \caption{Figure depicting a bicycle approaching a Class 2 observer with $\beta=0.9$ for frames 10 to 27. The left image in each frame is for the left aperture, likewise for the right.}
    \label{fig:leftbike}
\end{figure}

A human observer without equipment to detect light outside the visible range would therefore infer the appearance of the cyclist as it blocks the background they would otherwise see. They would, however, begin to notice the extreme elongation of the cyclist as it approaches, as visible from the side in Figure \ref{fig:deformedbike}. Figure \ref{fig:leftbike} also illustrates that, for large negative times, there is little difference in the distortions presented to both apertures, analytically illustrated by the convergence to 0 for $T\rightarrow\pm\infty$ in Figures \ref{fig:Phidiff} and \ref{fig:Thetadiff}. 

\begin{figure}[tbh]
    \centering
    \includegraphics[width=1\textwidth]{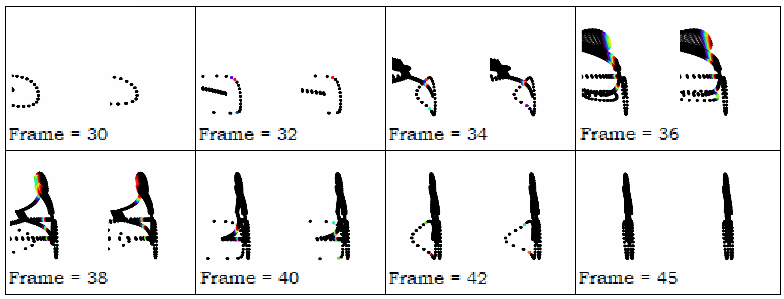}
    \caption{Figure depicting a bicycle abreast of a Class 2 observer with $\beta=0.9$ for frames 30 to 45.}
    \label{fig:middlebike}
\end{figure}

Around $T_{t'=0}$ in Figure \ref{fig:middlebike}, colours are shifted into the visual spectrum and enough radiation reaches the observer to be detected. In the case of an non-human observer, this information can be used to deduce first the form of the object, then its distance and velocity. It is also the point for which an observer notices the most deformation, as noted by Nowojewski \cite{Nowojewski2005}. For a human, a very thin strip of visibility would run vertically across the cyclist, revealing the distortions described by Mathews and Lakshmanan's transformations; namely, the front of the bicycle appears squashed and the left side of the cyclist's body appears rather horrifically twisted as it lags behind the right side. We also observe differences in the distortions presented to both apertures which, as analysed in the previous section, provide information about the object's motion and position.

\begin{figure}[tbh]
    \centering
    \includegraphics[width=1\textwidth]{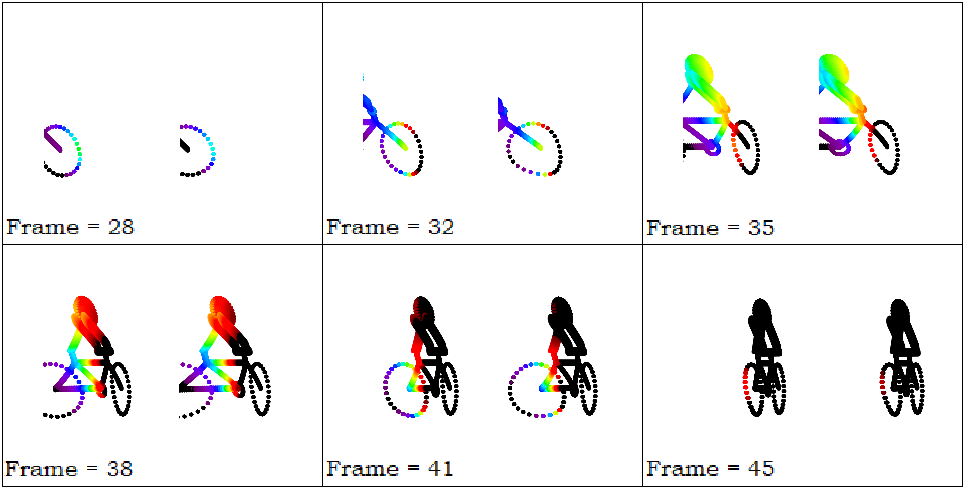}
    \caption{Figure depicting a bicycle abreast of a Class 2 observer with $\beta=0.6$ for frames 28 to 45.}
    \label{fig:middlebike0.6}
\end{figure}

For slightly slower objects (such as for $\beta=0.6$), the visual deformations are less marked but more of the bicycle is visible as the Doppler and intensity shifts remain within the visible spectrum. The same distortions, now with a thicker strip of visibility, are presented in Figure \ref{fig:middlebike0.6}. It is in the region of the point of closest approach for which the most striking distortions and the greatest differences between images presented to each aperture occur; it is from this disparity that the characteristic shapes of the plots produced in Section \ref{measuredistortions} arise, rendering it key for the determination of an object's distance and velocity. 

For $T>T_{t'=0}$, the light is again Doppler shifted out of the visual spectrum (into the ~1000nm short infrared) and the intensity arriving at the observer is minute as the majority is ``beamed" in front of the cyclist. We again represent the cyclist as a black solid since it would appear as an outline in front background radiation arriving from other sources, noticing the extreme contraction of the cyclist as the rear wheel appears to catch up with the front suddenly for times slightly greater than $T_{t'=0}$.

\begin{figure}[tbh]
    \centering
    \includegraphics[width=1\textwidth]{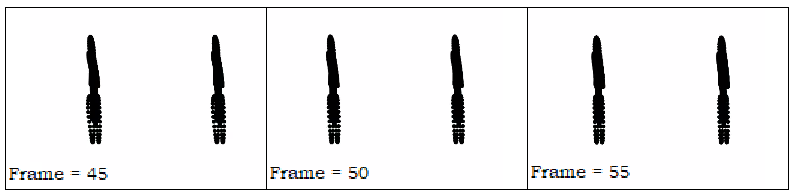}
    \caption{Figure depicting a bicycle receding from a Class 2 observer with $\beta=0.9$ for frames 45 to 55.}
    \label{fig:rightbike}
\end{figure}

\subsection{Fitting of Gamow's Cyclist}
\label{app:fittinggamowcyclist}

We now apply the methodology outlined in Section \ref{measuredistortions} to the case of Gamow's cyclist. By considering the forward-most pixel on the front wheel of the bicycle depicted in Figure \ref{fig:3Dbike}, we can produce similar azimuthal and polar plots to those in Section \ref{measuredistortions}, illustrating the viability of this method as a relativistic measuring tool. The simulated parameters are show in Table \ref{tab:sim}.

  \begin{table}[tbh]
    \centering
    \begin{tabular}{|c|c|}
         \hline
         Quantity & Value  \\
         \hline
         $\beta$  &  0.9  \\
         \hline
         $y_0$  &  3  \\
         \hline
         $z_0$  & 2  \\
         \hline
         $d$   & 0.1  \\
         \hline
    \end{tabular}
    \caption{Table of values utilised in creating the simulation. $y_0$ and $z_0$ are the coordinates of the forward-most pixel on the front wheel, $\beta$ is its velocity and $d$ is the aperture spacing.}
    \label{tab:sim}
\end{table}

\begin{figure}[tbh]
    \centering
    \includegraphics[width=1\textwidth]{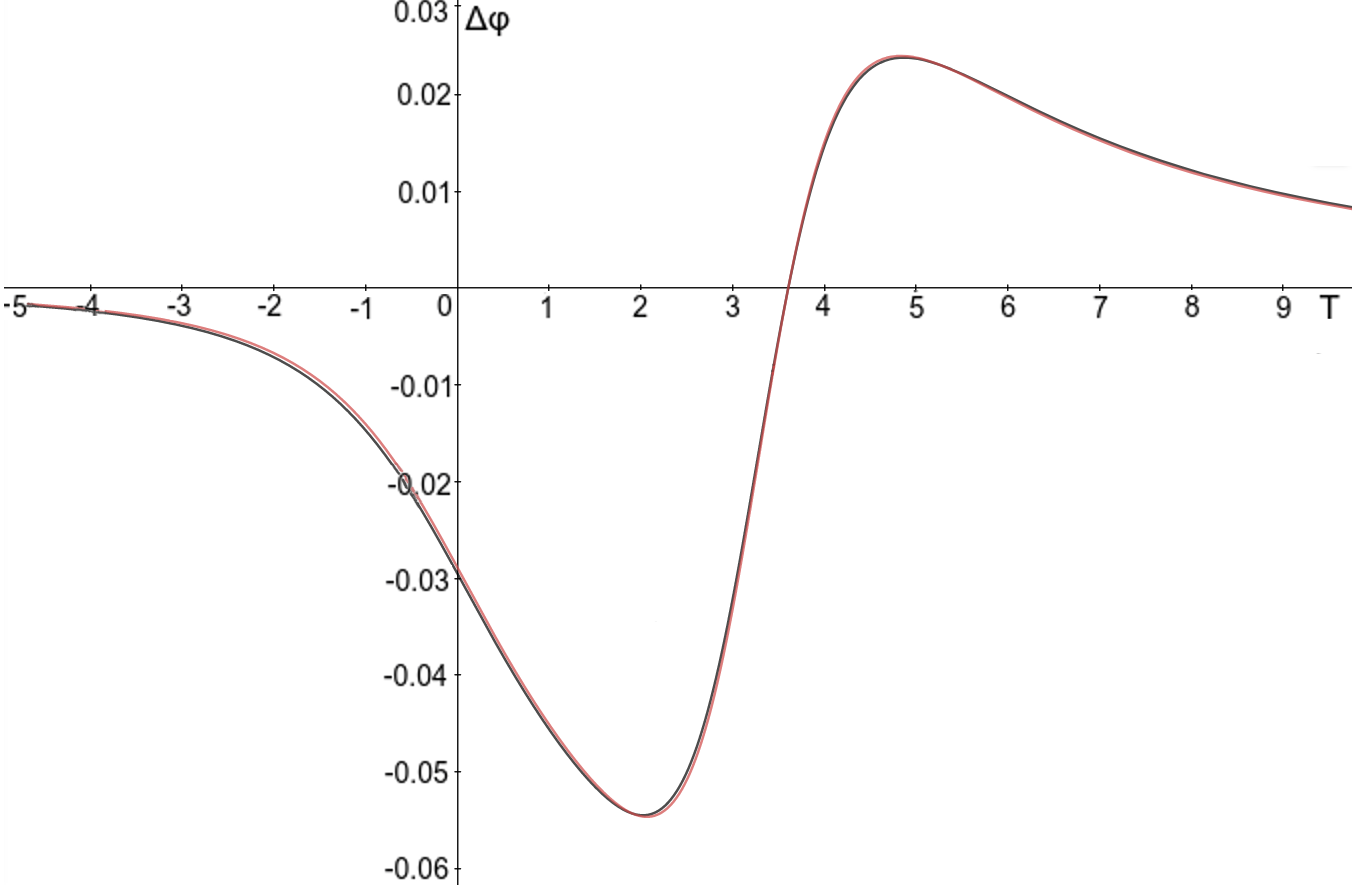}
    \caption{Figure illustrating the simulated azimuthal difference, $\Delta\phi$, and the fitted function. The simulation curve is in black and the fitting of Equation (\ref{eq:phidiff}) is in \textcolor{red}{red}.}
  \label{tab:fit}
\end{figure}

We present the success of fitting Equations (\ref{eq:phidiff}) and (\ref{eq:thetadiff}) to the simulation data using a non-linear bootstrapping regression method \cite{Oritin} in Table \ref{tab:fit}. The fitted quantities for the azimuthal difference, $\Delta\phi$, are given in Table \ref{tab:azi}

\begin{table}[tbh]
    \centering
    \begin{tabular}{|c|c|c|}
         \hline
         Quantity & Value & Uncertainty \\
         \hline
         $\beta$  &  0.901 & 0.005 \\
         \hline
         $y_0$  &   2.99 & 0.05 \\
         \hline
         $z_0$  & 2.02   &   0.07 \\
         \hline
    \end{tabular}
    \caption{Table of values and uncertainties obtained from fitting of the azimuthal angle difference for a Class 2 observer with $d=0.1$.}
  \label{tab:azi}
\end{table}

\begin{figure}[tbh]
    \centering
    \includegraphics[width=1\textwidth]{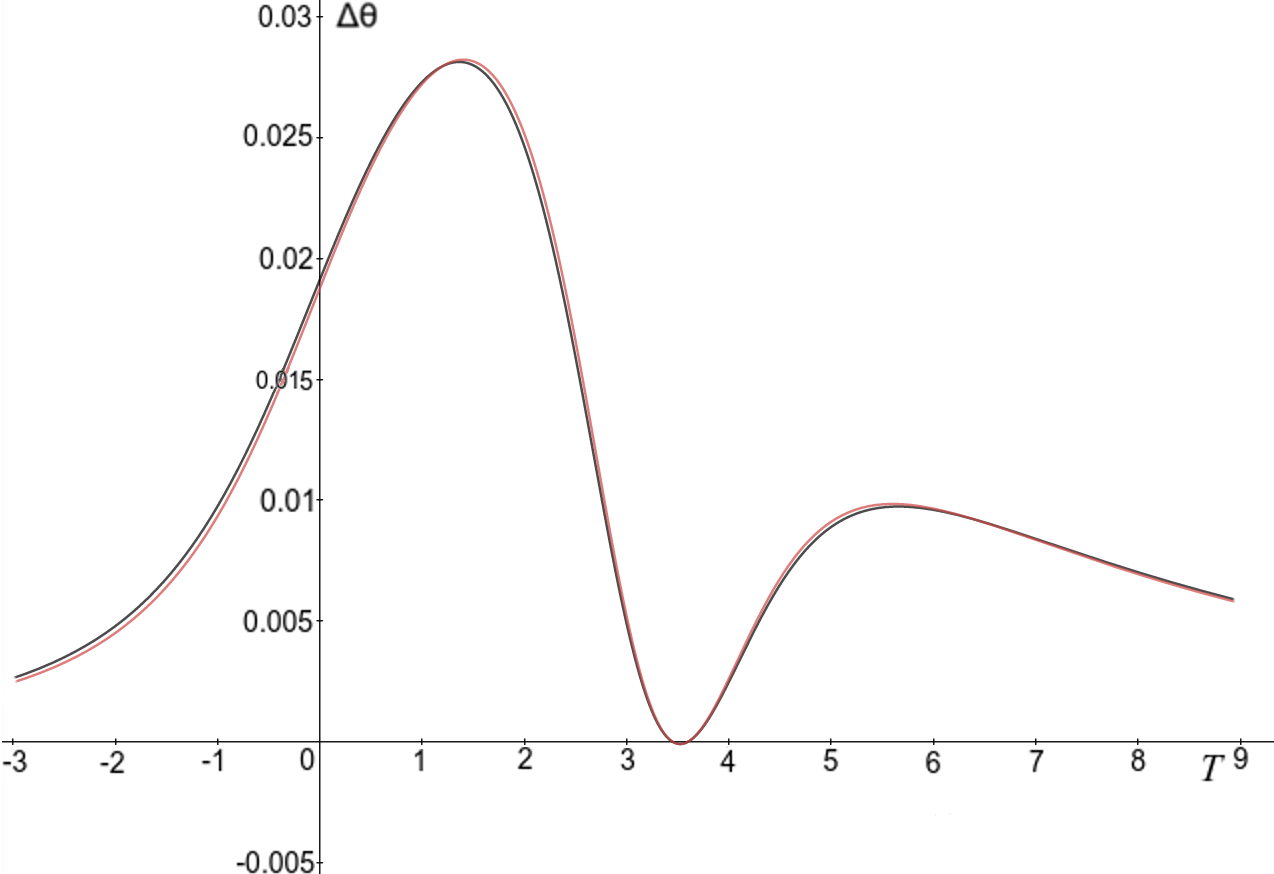}
    \caption{Figure illustrating the simulated polar difference, $\Delta\theta$, and the fitted function. The simulation curve is in black and the fitting of Equation (\ref{eq:thetadiff}) is in \textcolor{red}{red}.}
    \label{fig:Thetafitting}
\end{figure}

The complimentary fitting for the polar angle is presented in Figure \ref{fig:Thetafitting}. We observe larger uncertainties in the fitted values as the two intersections with the x-axis approach the same value; as the smaller-$T$ intersection is $\beta$-dependent and the one at larger-$T$ is not, the close proximity of the two increases the fitted uncertainties.  The fitted values for the polar difference, $\Delta\theta$, are shown in Table \ref{tab:pol}.

\begin{table}[tbh]
    \centering
    \begin{tabular}{|c|c|c|}
         \hline
         Quantity & Value & Uncertainty \\
         \hline
         $\beta$  &  0.90 & 0.01 \\
         \hline
         $y_0$  &   3.02 & 0.06 \\
         \hline
         $z_0$  & 1.99   &   0.09 \\
         \hline
    \end{tabular}
    \caption{Table of values and uncertainties obtained from fitting of the polar angle difference for a Class 2 observer with $d=0.1$.}
  \label{tab:pol}
\end{table}

Uncertainties in the relativistic parameters fitted were calculated using a bootstrapping method, shown to be a better indicator of uncertainty in non-linear fitted parameters than comparable Monte Carlo methods or linearisation methods \cite{Hu2015}.

Averaging these quantities for azimuthal and polar fitting with weight determined by the uncertainty and propagating error diferentially, we obtain $\beta=0.90\pm0.01$, $y_0=3.00\pm0.05$ and $z_0=2.00\pm0.08$, the exact values of the cyclist created for the simulation detailed in Figure \ref{fig:3Dbike}. The errors in these quantities of course do not reflect any experimental uncertainty as they were obtained from a deterministic simulation; they do however represent the errors associated with the fitting of data of this form with Equations (\ref{eq:phidiff}) and (\ref{eq:thetadiff}) using a non-linear regression method \cite{Origin}. It is thus suggested that this methodology has potential as a tool for the accurate measurement of relativistic objects, determining both their objective velocity and position, using Class 2 observers. These quantities can then be used to transform relativistic photographs and spectral measurements back to their rest forms, a more useful format for physical analysis.
\\
\\
With regard to applications, it is posited by the authors that an interferometer using this fitting technique could be either implemented as part of ground based arrays such as those used in aperture synthesis \cite{Jennison1958} or as a relativistic probe capable of probing astronomical bodies while travelling at great speed \cite{Christian2017}. In the case of the former, inverse Fourier transforms are utilised to resolve the image for multiple observers; as such, an extension to this work would be to generalise the solutions for Class $N$ observers and implement differential geometry for a smoothly varying time difference, $\Delta T$, across all observers in the aperture that could be used in differential calculus. In the case of the latter, further work must be carried out to generalise the solutions for General Relativistic effects and implement solutions to the geodesic equations for photon trajectories to allow for inclusion within Christian and Loeb's suggested probe which accounts for gravitational effects.

\section{Concluding Remarks}
\label{conclusion}

In summary, the visual appearance of Gamow's cyclist has been comprehensively re-analysed in 3-dimensions, accounting for radiative shifts as well as apparent distortions in its shape (which comprise a non-linear shear and elongation/contraction depending on position), and for a binocular observer. The actual appearance of the cyclist for a human observer is then speculated, taking into account the difference in distortions presented to both eyes as well as the limits of the visual spectrum. A method of determining the position and velocity of moving objects is also illustrated, making use of the azimuthal and polar disparities between apertures, providing an estimate of the uncertainty by comparing the fitting of both curves. This methodology is then applied to the simulation of Gamow's relativistic cyclist, correctly reproducing the simulated parameters with the uncertainties obtained from non-linear fitting. Finally, it is suggested that these methods could be integrated within a relativistic probe \cite{Christian2017} to observe stationary objects as they move past, or a synthetic aperture which would allow the probing of distant, relativistic objects.

\enlargethispage{20pt}

\ethics{There are no ethical issues associated with this research}

\dataccess{This article has no additional data.}

\aucontribute{EC-J performed the calcualtions associated with this project.  Both authors discussed the ideas involved, were both involved in writing the paper, and agreed the final submitted form.}

\competing{The authors declare no competing interests.}

\funding{PDS acknowledges support from UK STFC grant ST/P005314/1}

\ack{The authors thank Dr Maxime Delorme for help in constructing the cyclist model in Python}

\appendix
\section{Full expression for Right-Aperture Apparent Position}
\label{app:fullexpression}

A shortened expression for the apparent position of a point for the right aperture was presented in Equation (\ref{eq:righteyeposition}) in the interests of preserving space where the non-linearities in $T$ have been suppressed within the $\Delta T$ term. Here, we give the entire, lengthy equation only in terms of $T$.  Note that $c=1$ has been used in this equation for (attempted) brevity:
\begin{align*}
    x'  &= \gamma^2\Vast[x_0+\beta\Vast\{T+\Vast(\sqrt{y_0^2+z_0^2+\left(\beta\left(\gamma^2\left(T+x_0\beta\right)-\gamma\sqrt{\gamma^2\left(\beta T+x_0\right)^2+y_0^2+z_0^2}\right)+\frac{d}{2}\right)^2} \\
     &- \sqrt{y_0^2+z_0^2+\left(\beta\left(\gamma^2\left(T+x_0\beta\right)-\gamma\sqrt{\gamma^2\left(x_0+\beta T\right)^2+y_0^2+z_0^2}\right)-\frac{d}{2}\right)^2}\Vast)\Vast\}\Vast] \\
     &- \beta \gamma\Vast[\gamma^2\Vast\{x_0+\beta\Vast(T+\Vast(\sqrt{y_0^2+z_0^2+\left(\beta\left(\gamma^2\left(T+x_0\beta\right)-\gamma\sqrt{\gamma^2\left(x_0+T\beta\right)^2+y_0^2+z_0^2}\right)+\frac{d}{2}\right)^2} \\
     &- \sqrt{y_0^2+z_0^2+\left(\beta\left(\gamma^2\left(T+x_0\beta\right)-\gamma\sqrt{\gamma^2\left(x_0+\beta T\right)^2+y_0^2+z_0^2}\right)-\frac{d}{2}\right)^2}\Vast)\Vast)\Vast\}^2+y_0^2+z_0^2\Vast]^{\frac{1}{2}} \\
\end{align*}

\bibliographystyle{RS}
\bibliography{refs}

\end{document}